\documentclass[]{osa-article}

\journal{oe}
\articletype{Research Article}

\usepackage{siunitx}

\usepackage{subfigure}
\usepackage{tikz}
\usetikzlibrary{decorations.pathmorphing}
\usetikzlibrary{decorations.markings}
\usetikzlibrary{positioning}
\usetikzlibrary{patterns}
\usetikzlibrary{arrows}
\tikzset{cross/.style={cross out, draw, 
         minimum size=2*(#1-\pgflinewidth), 
         inner sep=0pt, outer sep=0pt}}
\usepackage{siunitx}
\usepackage{multirow}
\usepackage{hhline}
\usepackage{acronym}
\usepackage{orcidlink}
\definecolor{orcidlogocol}{HTML}{A6CE39}

\acrodef{FIOS}[FIOS]{Fiber Injector Optical Sub-assembly}
\acrodef{LISA}[LISA]{Laser Interferometer Space Antenna }
\acrodefplural{FIOS}[FIOS]{Fiber Injector Optical Sub-assemblies}
\acrodef{DBB}[DBB]{Diagnostic Breadboard}
\acrodef{QPD}[QPD]{Quandrant Photodiode}
\acrodef{QMFC}[QMFC]{Quasi Monolithic Fiber Collimators}
\begin{document}

\title{Quasi Monolithic Fiber Collimators}

\author{Jonathan Joseph Carter\orcidlink{0000-0001-8845-0900},\authormark{1,2,*}  Steffen Böhme,\authormark{3} Kevin Weber\orcidlink{0009-0003-5199-8712}, \authormark{1,2} Nina Bode,\authormark{1,2}, Karina Jorke\authormark{3}, Anja Grobecker \authormark{3}, Tobias Koch\authormark{3}, Simone Fabian\authormark{3}, and Sina Maria Koehlenbeck\orcidlink{0000-0002-3842-9051}\authormark{1,2}}

\address{\authormark{1}Max Planck Institute for Gravitational Physics (Albert Einstein Institute), Callinstraße 38, 30167 Hannover, Germany\\
\authormark{2}Institut für Gravitationsphysik der Leibniz Universität Hannover, Callinstraße 38, 30167 Hannover, Germany\\
\authormark{3}Fraunhofer Institute for Applied Optics and Precision Engineering, Albert-Einstein-Str. 7, 07745 Jena, Germany}

\email{\authormark{*}jonathan.carter@aei.mpg.de} 
\begin{abstract}
    Interferometric displacement measurements, especially in space interferometry applications, face challenges from thermal expansion. Bonded assemblies of ultra-low thermal expansion glass-ceramics offer a solution; however, transitioning from light transport in fibers to free beam propagation presents a notable challenge. These experiments often need an interface to convert between laser beams propagating through fiber optics into a well-defined free beam and vice versa. These interfaces must also be made of rigid glass pieces that can be bonded to a glass base plate. Current designs for these fiber collimators, often called fiber injector optical sub-assemblies, require multiple glass parts fabricated to very tight tolerances and assembled with special alignment tools. We present a simplified quasi-monolithic fiber collimator that can generate a well-collimated laser beam. The complexity and tolerances of bonding are reduced by combining the alignment of the fiber mode to the imaging lens in one step with active mode control: the welding of the fiber to the glass body. We produce several of these designs and test that the desired light field is achieved, its profile is described as a Gaussian beam, and the beam-pointing stability is acceptable for such a piece. In each case, they perform at least as well as a standard commercial fiber collimator. These Quasi Monolithic Fiber Collimators offer a promising and easy-to-implement solution to convert between free beam and fiber-coupled lasers in experiments sensitive to long term thermal drifts. 
\end{abstract}

\section{Introduction}
Displacement measurements with interferometers have become increasingly sensitive to the extent that the thermal expansion of individual components can dominate the phase measurement. The drifts induced by thermal expansion can cause many unwanted effects in laser interferometry, such as a reduction in optical contrast from interference, less efficient coupling to specific components, or direct changes to beam path length, such as through tilt-to-length coupling.  Space interferometry, in particular, has very high demands on the thermal stability of earth-based pre-experiments. The problem is reduced using bonded assemblies of ultra-low thermally expanding glass ceramics. The transition of light transport from fibers to free beam propagation is not an exception but a particular challenge. These interfaces use an aspheric lens held at a specific distance along a beam path. Often, the desired beam is optimally collimated, maximizing the Rayleigh range of the beam.
\par
The beam pointing and shape resulting from this transition must be well controlled. To achieve a stable beam, the fiber must be well centered on the lens in the directions transverse to the beam propagation and held at a specific position along the direction of travel. Due to the added complexity, active control of the alignment is not a desired solution in space missions. Hence, to maintain alignment, these pieces should be bonded in a sub-assembly for applications sensitive to thermal drifts. 
\par
Space Missions using laser interferometry overcome this using \acp{FIOS}. \cite{Killow2016,Chwalla2016}.  The LISA pathfinder \ac{FIOS} \cite{Killow2016} consisted of several parts bonded in series to a baseplate made of Heraeus Suprasil 1 Fused Silica. Hydroxide Catalysis bonding was used to bond most of the components together \cite{Veggel2014,Bogenstahl2017}. The space-qualified design survived all the requisite tests for such a mission. In particular, there was a need for the component to withstand larger thermal cycles and mechanical force without changing the beam pointing significantly. Designs used in a \ac{LISA} pre-flight optical bench \cite{Chwalla2016} used an additional fused silica spacer on the fiber to minimize power density at the fiber tip, making it less susceptible to contaminants in the lab. Furthermore, the spacer produces additional thermal stability in the thermal conditions in most labs. These preflight test beds must also have specific beam profiles. 
The precision of these devices' resulting modes is defined by how precisely the bonding stages can be performed. 
The Grace Follow on laser ranging interferometer's \cite{Danzmann2017,nicklaus2017} optical bench also required the development of \ac{FIOS}. Again, the focus of these pieces was to survive vibrational shocks and thermal cycling. These FIOS, however, were the first to show a fiber welding process that could directly attach the fiber to a spacer \cite{Boehme2009}, which is the method used to attach the fiber to the cuboid in this work. 
\par
We present a novel approach to making pre-flight test \ac{FIOS}, with a simplified design bonded without a baseplate, with the lens and fiber directly attached to a spacer. The spacer is the length between the fiber and lens, which defines the resultant mode. Therefore, the tolerance on the spacer length controls the tolerance on the resulting mode. The lens axis and the fiber bonded in place with  CO$_2$ lasering weld after an active alignment process, which achieves a sub-\SI {}{\micro\meter} precision on the alignment. This approach means the alignment and tolerance are no longer defined by the skill of the group bonding the device together but by the manufacturing tolerances of the parts, which can be much better controlled between different setups.

\begin{figure}
    \centering
    \begin{tikzpicture}[>=latex,scale=.7,decoration=snake,decoration=snake,every node/.style={scale=.8}]

   \node(pic)at(-3.5,2.3){\scalebox{-1}[1]{\includegraphics[width=.5\textwidth]{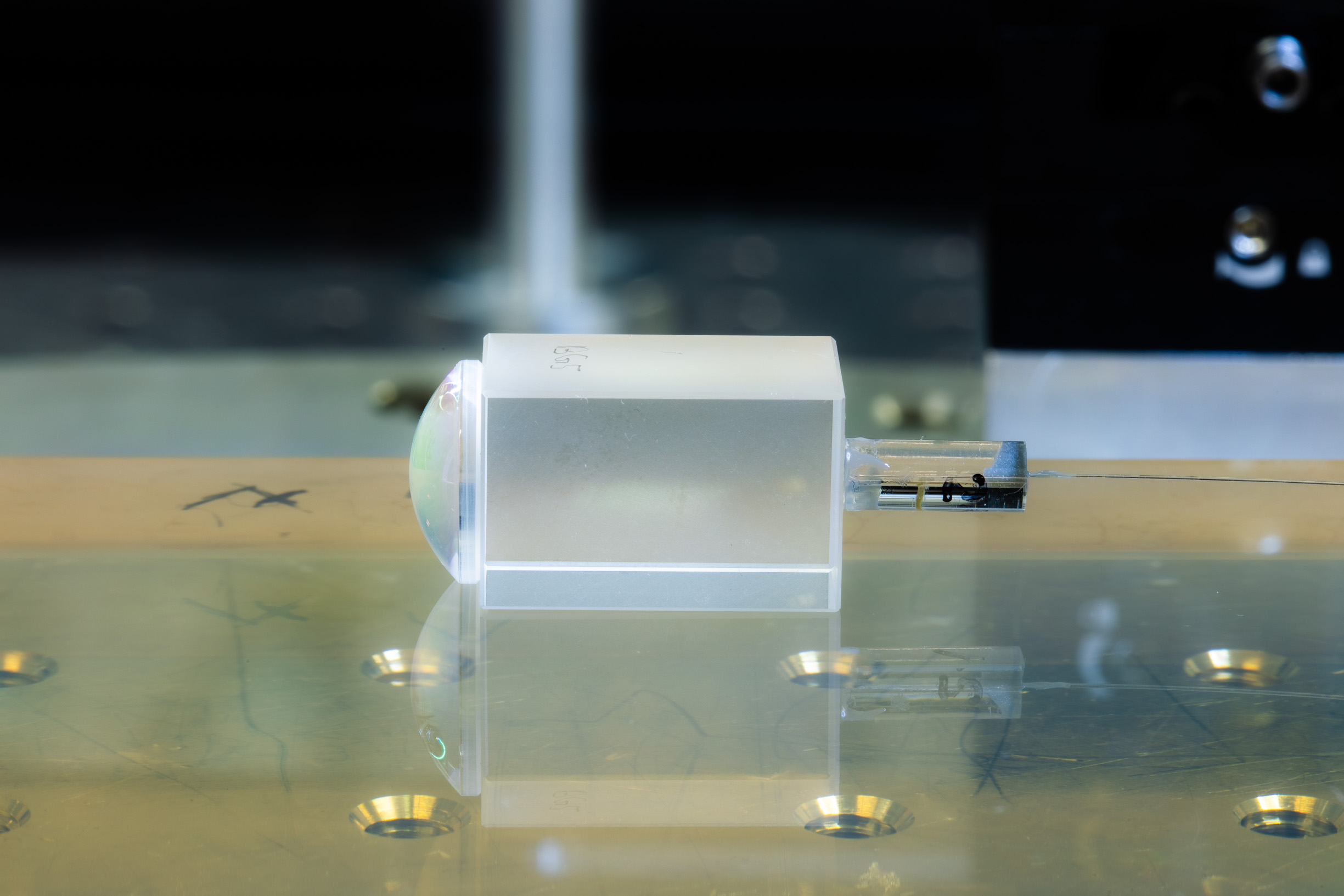}} };
    \draw[-triangle 45,](-6.5,-.5)--(-5.,2.2);
    \node[below]at(-6.5,-.5){{Ferrule for Support}};

    \draw[-triangle 45,](-7.5,1.7)--(-6.7,2.2);
    \node[left]at(-7.5,1.7){{Optical Fiber}};

    \draw[-triangle 45](-3.5,-.5)--(-3.5,2);
    \node[below]at(-3.5,-.5){{Glass Cuboid}};

    \draw[-triangle 45](-.5,-.5)--(-2.,2);
    \node[below]at(-.5,-.5){{Aspheric Lens}};

    \draw[-triangle 45](-8,3.5)--(-4.7,2.4);
    \node[left,align=right]at(-8,3.5){Fibre Weld\\ Region};
    \draw[-triangle 45](.5,3)--(-2.5,2.5);
    \node[right,align=left]at(.5,3){Bonded Surface\\ (Optical Contact \\ or Hyroxide Catalysis\\ Bond)};
    
    \end{tikzpicture}

    \caption{A photo of the Quasi Monolithic Fiber Coupler. The device consists of three parts, the fiber, the cuboid, and the aspheric lens, each colored differently. The interfaces between each part are one key aspect of the design. The fibre is fused directly to the glass cuboid with a fibre welding process, the lens was attached with either a optical contact, or a hydroxide catalysis bond. }
    \label{fig:tiltedcav}
\end{figure}
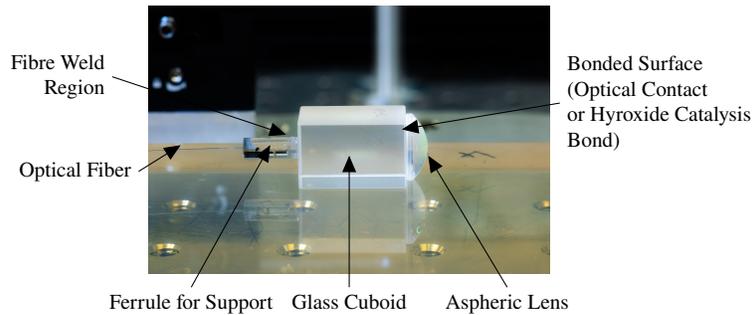

\section{Quasi Monolithic Fiber Collimators}
\label{sec:design}
A Quasi Monolithic Fiber Coupler that collimates a beam must take the beam from the single mode polarisation maintaining fiber, with a mode field diameter of several \SI{}{\micro\meter}, and convert it to a beam with sufficient Rayleigh range suitable for tabletop experiments, typically at least a few meters. This beam must have a transmitted profile of the correct shape and not be deformed. The whole assembly must limit optical back reflections into the fiber. Furthermore, the resulting beam must be rigid enough to minimize beam pointing noise.
\par
The fiber collimator's design is optimized for these issues. Its design is shown in Figure \ref{fig:tiltedcav}. The design has three components: the fibre injects light into a glass cuboid, whose length allows the appropriate expansion of the light so that the final piece, an aspheric lens, can optimally collimate the beam. The piece being rigidly bonded together minimizes pointing noise. As the beam propagates only through fused silica, we minimize backscattering induced by refractive index changes. Stock aspheric lenses and fibers are used, while the spacer length is tuned to provide an optimally collimated beam.  
\par
As a rigid assembly is desired, the aspheric lens needed to be attached to the glass cuboid through a bonding method. We wanted no refractive index changes between the different interfaces to minimise back reflection into the fiber. Two different bonding methods were explored for this. The first was a simple optical contact. A large force is applied to the lens to push it onto the glass cuboid. Van der Waals forces then hold the lens in place. The strength of this bond depends highly on the surface quality of the two optics. It was found here that with the stock aspheric lens surface quality of 300\,nm RMS and a fine ground polish on the glass cuboid lead to a successful optical contact on most samples. 
\par
In cases where a higher bond strength is needed, an alternative bond using hydroxide catalysis bonding(also known as silicate bonding) was compared  \cite{Veggel2014}. Here, a suitable reagent, such as potassium hydroxide or sodium hydroxide\cite{Veggel2014}, is applied to each surface, which causes long siloxane chains to form. When two surfaces treated this way are brought into contact, the chains intertwine, and a permanent, strong bond is formed.The bond makes the assembly significantly more shock-resistant. As silicate bonding has less stringent requirements on surface roughness, it can also be used when surface roughness or planarity is insufficient for optical contact. 
\par
\begin{figure}
    \centering
    \includegraphics[width=.9\textwidth]{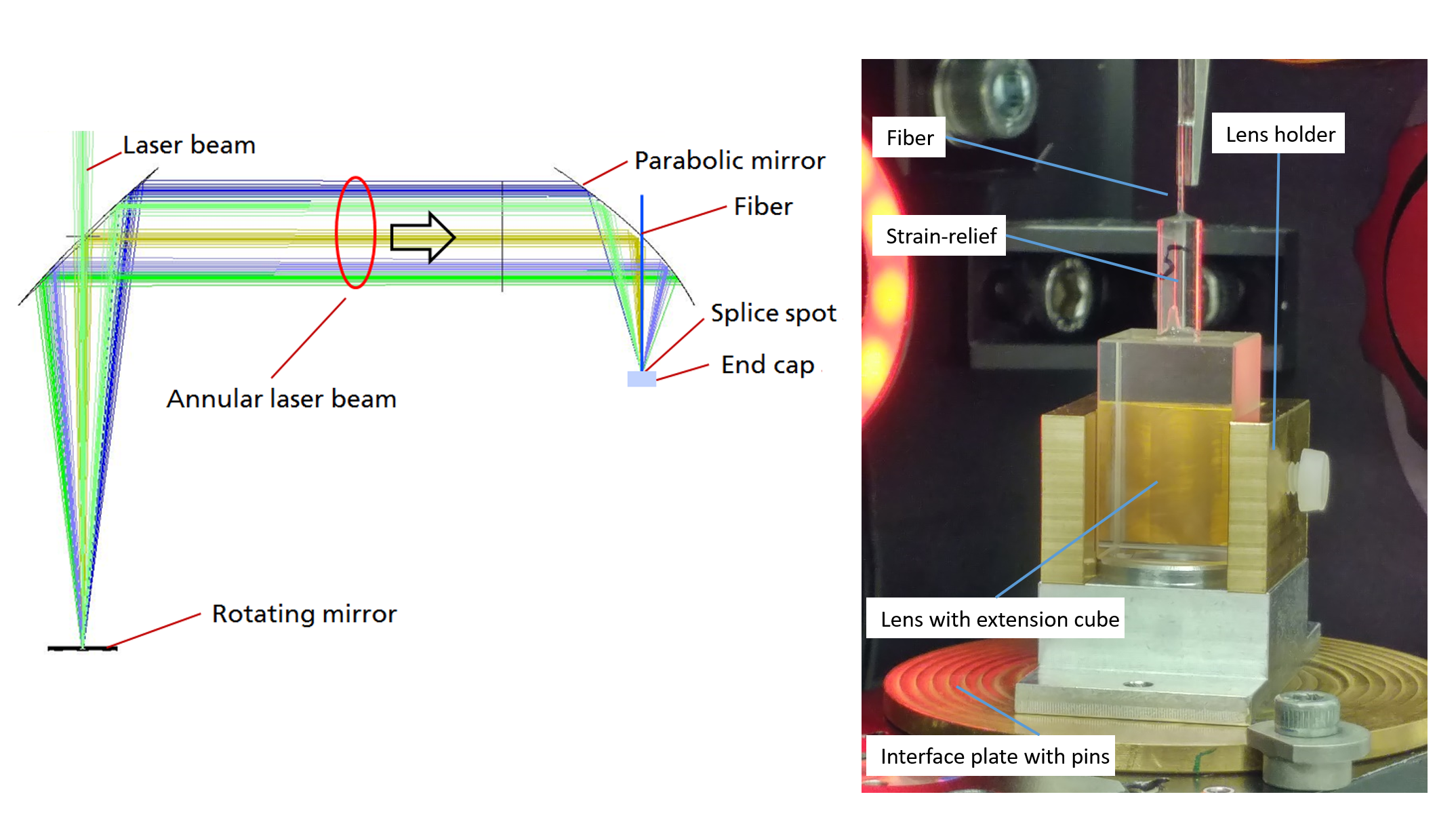}
    \caption{A figure showing how the fiber is welded onto the cuboid. A laser is shone onto a rotating mirror, creating an annular laser beam. In turn, this is focused on the precise location of the fiber weld, which creates a heated and therefore fused region. A photo of a collimator being fused is shown in the right panel. }
    \label{fig:FSM}
\end{figure}

The light travels from the fiber directly into the glass cuboid. The fiber was fused directly to the cuboid using a fiber welding process. Welding the optical fiber requires the cuboid to be bonded to the selected aspherical lens first. Using a Shack Hartmann wavefront sensor, the optical axis of the asphere can then be aligned to the optical fiber axis with sub-\SI{}{\micro\meter} precision by active alignment. In addition, one of the outer flat surfaces of the cuboid is aligned with the polarisation plane of the fiber. The fiber is welded to the glass cuboid with a CO$_2$ laser-based device \cite{Boehme2017}. Figure \ref{fig:FSM} shows this process. During the welding process, the two surfaces are heated and then the previously aligned fiber is placed on the cuboid surface and fused to it. If the parameters of the CO$_2$ laser-based welding process are selected appropriately, the beam quality of the fiber-cuboid-aspheric lens assembly does not change significantly after welding. This has already been successfully demonstrated in previous space projects Grace-Follow On \cite{Boehme2009} and Methane Remote Lidar Mission (MERLIN) \cite{Pierangelo2016} with specially designed aspherical rod lenses. At the location of the fiber weld, a strain relief was included to provide additional strength to support the joint.
\par
\subsection{Parameter tuning}
Depending on what pieces in the collimator are fixed, every other parameter will have a definitive optimum value. These optimum values must be calculated. This is best done using a ray transfer matrix 
\begin{equation}
     \begin{bmatrix} q_{\rm{col}} \\ 1 \end{bmatrix}=k\begin{bmatrix} A & B \\ C & D \end{bmatrix}\begin{bmatrix} q_{\rm{fib}} \\ 1 \end{bmatrix},
\end{equation}
where $k$ is a normalisation factor, $q$ indicates the complex beam parameter
\begin{equation}
    q=(z-z_0)+iz_{\rm{R}},
\end{equation}
with a Rayleigh range
\begin{equation}
    z_{\rm{R}}=\dfrac{\pi {\omega_0}^2n_2}{\lambda},
\end{equation}
where $\omega_0$ is the beam radius.
The suffixes define whether the beam is the collimated, free beam or the beam in the fibre.
The $ABCD$ matrix is the matrix of the cuboid and lens assembly given by 
\begin{equation}
    \begin{bmatrix} A & B \\ C & D \end{bmatrix}=\begin{bmatrix} 1 & 0 \\ \dfrac{n_1-n_2}{R_{\rm{lens}}n_2} & \dfrac{n_1}{n_2} \end{bmatrix}
\end{equation}
where $n_1$ indicates the refractive index for the lens material, $n_2$ that of the propagating free beam, and $R_{\rm{lens}}$ the radius of curvature of the lens surface. For fused silica optics and 1064\,nm laser light, we estimate $n_1$ to be 1.45 and assume the resulting beam propagates in air such that $n_2\sim$1. We now have a constrained set of parameters to study; the mode field diameter (MFD) of the fiber, the position of the lens with respect to the fiber, equivalent to cuboid length, and the lens' radius of curvature. This analysis treats $z=0$ as the interface between the lens and air. This effectively makes $z_0$ of $q_{\rm{fib}}$ equal to lens+cuboid length, $L$. 
\par
\begin{figure}
    \centering
    \includegraphics[scale=.65] {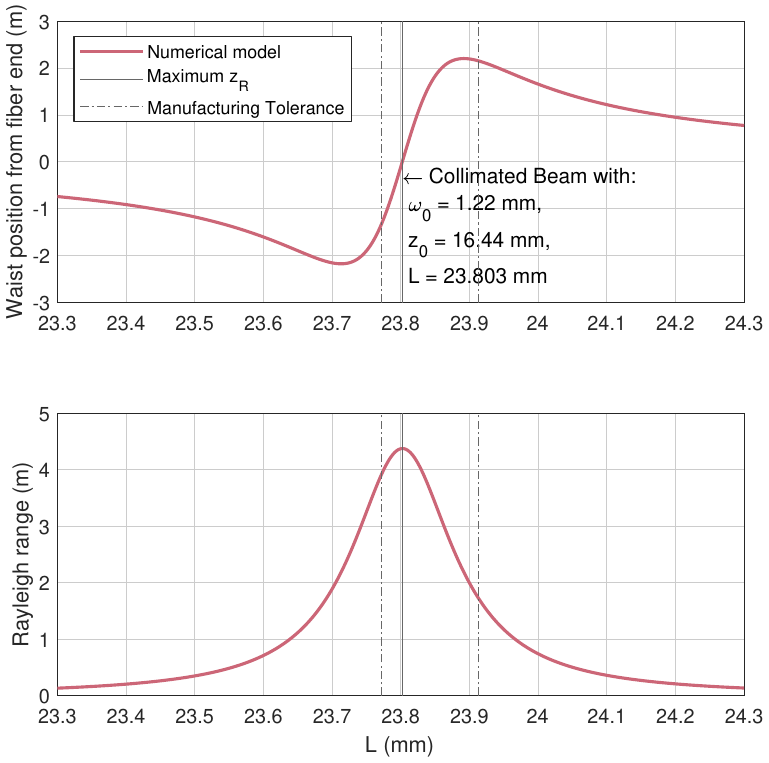}
    \caption{Example optimisation of $L$ (the length of the cuboid and lens) for an example fiber/lens pair. This specific case considered a MFD of \SI{5.5}{\micro\meter} and lens $R_{\rm{lens}}$ of 7.387\,mm. The optimally collimated beam occurs when $z_{\rm{R}}$ is maximised. The optimum has been highlighted with a solid line. The dashed lines show the range of parameters possible from tolerance on $L$.}
    \label{fig:lensLength}
\end{figure}

The best collimation on a beam is achieved when $z_{\rm{R}}$ is maximised. We show a specific case of a fiber with a MFD of \SI{5.5}{\micro\meter} and $R_{\rm{lens}}$ of 7.387\,mm.  The optimally collimated beam is shown in Figure \ref{fig:lensLength} by the dashed line. Here, the response of the Rayleigh range to length changes is shown to be relatively shallow around the peak, but the waist position will change dramatically with respect to small length changes. 

\begin{figure}
    \centering
    \subfigure[MFD]{\includegraphics[width=.49\textwidth] {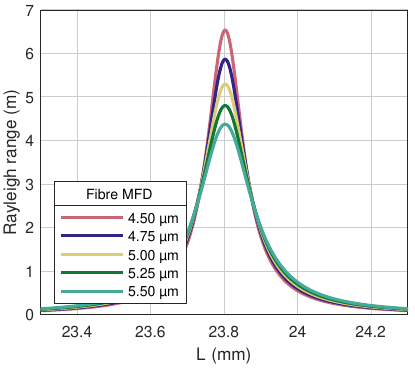}}
     \subfigure[$R_{\rm{lens}}$]{\includegraphics[width=.49\textwidth] {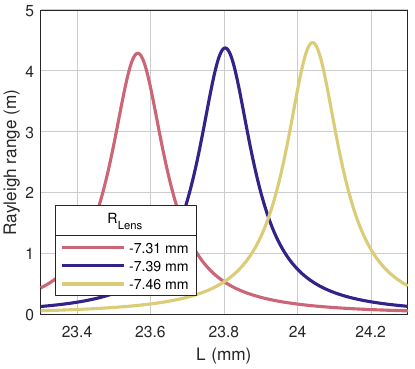}}
    \caption{The Rayleigh range of the collimator evolves with respect to length for different $L$. a) shows the case for different MFDs, where a larger fiber MFD results in a small Rayleigh range being possible, but ultimately will not change the need of $L$. b) show the case for different $R_{\rm{lens}}$, and how both a very sharp response is seen to small changes in the value. }
    \label{fig:errorOfmflensx}
\end{figure}
Along with uncertainties on $L$, there are corresponding uncertainties on the MFD of the fiber, and $R_{\rm{lens}}$. We explore these effects in Figure \ref{fig:errorOfmflensx}. Fiber MFD diameters typically have a broad tolerance on quoted values of $\sim\pm$\SI{1}{\micro\meter}. The variation in MFD does not affect the optimal value of $L$. It does, however, place a maximum limit on the resultant $z_{\rm{R}}$. Fibers should, therefore, be chosen with small MFD to best optimise the range. Small changes in $R_{\rm{lens}}$ cause significant shifts in ideal values of $L$. Hence, a very accurate measurement of the radius of $R_{\rm{lens}}$ is needed before the cuboid is made or some means of tuning the cuboid length to match the appropriate lens diameter.   
\subsection{Manufacture of Parts}
A total of eight samples were produced. The samples were produced using slightly different methods and parts, due to the desire to test methods and the availability of parts. The first five samples were produced using an optical contact on the lens to the cuboid unmodified. The later three samples were polished to reduce their length and best match the ideal beam, but due to a mismeasurement in overall length, they were polished too short. Four samples used a hydroxide catalysis bond to attach the lens to the cuboid instead of optical contacting. This allowed a direct comparison of the two methods of bonding.
\par
The aspheric lens used were stock parts from Aspericon. The pieces had a thickness of \SI[separate-uncertainty = true]{4000(50)}{\micro\meter}, were made of Corning 7890-0F fused silica, and had an effective focal length quoted as 15.0\,mm.  To provide the ideally collimated beam, the cuboid had to be \,19.80\,mm long.  This length was deliberately tuned to be slightly longer than this (19.84\,mm), so that they could be reduced to the correct length, if needed, by a more precise laser ablation process. The cuboid also made of Corning 7890-0F Fused Silica Quartz. Ultimately, this step was never tested. The best tolerance that could be achieved on this overall length through standard manufacturing techniques, without supplementary polishing or laser ablation, was \SI{20}{\micro\meter}. The regions bounded by the dashed lines in Figure \ref{fig:lensLength} show the range of resulting beams possible due to this tolerance. If both the lens and cuboid were at the upper limit, then a rayleigh range of 1.7\,m could still be achieved, although the $z_0$ could occur at any point in its possible range. 
\section{Evaluation of Performance}
\label{sec:perf}
The mode matching of the samples could be tested using a relatively simple optical setup where beam profiles were measured along the direction of propagation through the air. The mode purity and pointing stability were tested using an Albert Einstein Institute's \ac{DBB}. The \ac{DBB}'s function is thoroughly described by Kwee et al. \cite{Kwee2007}. In Short, the experiment uses a ring cavity to break down the test beam into its constituent modes, and the relative beam pointing using two \acp{QPD} in the reflection port of the cavity. Throughout the section, we compare when relevant to a commercially available fiber collimator from Schäfter + Kirchoff GmbH, a standard collimator used in many high-precision experiments that do not use fixed ULE breadboards and bonded optics. This acts as a reference for the beam properties one typically needs for high-precision physics experiments.
\subsection{Mode matching}
As the aim of the collimator was to achieve a specific, well-collimated beam profile, we needed to measure the profile of each sample produced. This was done by using a beam profile camera to take pictures of the beam over a length of 2\,m, with photos taken every 10\,cm. The measured beam sizes over space were fitted to the standard equation of a Gaussian beam. 
The first four samples produced and tested had a $\omega_0$ of \SI[separate-uncertainty = true]{1220(50)}{\micro\meter}, with a $z_0$ of \SI[separate-uncertainty = true]{250(20)}{\milli\meter}. There was very slight astigmatism on each sample, with a difference between the major and minor axis in $\omega_0$ of \SI[separate-uncertainty = true]{50(25)}{\micro\meter}.
\par
As all the lenses' of the initial batch were too long, it was planned to try a tuning method of the length. The lens was first to be polished slightly, within \SI{20}{\micro\meter} of the exact required length, then use a C0$_2$ laser ablation process to tune the lengths during the fiber welding process precisely. Unfortunately, the pieces were slightly overpolished during the polishing step, so they were too short before the ablation began; hence, this step was not tested. After the polishing step, the pieces had an $\omega_0$ of \SI[separate-uncertainty = true]{1100(110)}{\micro\meter}, and $z_0$ of \SI[separate-uncertainty = true]{-230(50)}{\milli\meter}, suggesting the overpolish was about the length of the original error. 
\par
The $\rm{M}^2$ of 4 of the first 4 pieces was measured. These were estimated using the beam profile over the length of 5\,m with samples every 50\,cm. The resulting beam was then fitted to
\begin{equation}
    \omega^2(z)=\omega_0^2+\mathrm{M} ^4\left(\frac{\lambda}{\pi \omega_0}\right)^2\left(z-z_0\right)^2
\end{equation}
with a linear regression tool. The samples' mean $\rm{M}^2$ was 1.28, but with a considerable range (1.125-1.474). Still, the resulting profiles from these collimators have excellent quality and are suitable for most interferometric purposes.
\par
Regardless, all samples showed Rayleigh ranges in the order of metres, which is more than sufficient for most tabletop applications. Further developments in fine-tuning specific beams, such as mode-matching optical resonators or precise recoupling to fibers in designated places, are required. Ablation techniques should be further developed to this end to allow for fine-tuning of length for such purposes. However, for most applications, these are unnecessary steps for fiber injectors, and even relatively simple techniques can achieve beams sufficiently collimated for most interferometric purposes.

\subsection{Mode purity}
A beam shape well-defined by a Gaussian Mode is needed for most applications. Deformations in beam shape can lead to excess noise, scattered light, and, when an optical resonator is used, a direct loss in circulating power. We therefore study the overall contamination of the beam with higher-order Hermite-Gaussian modes (HG(m,n)). These higher-order modes can be used to estimate the total power lost from the fundamental mode, and the total power in these modes tells us how pure the mode of the beam is. In addition, by studying the contents of these higher-order modes, we can comment on the actual causes of the beam deformation. 
\begin{figure}
    \centering
    \includegraphics[width=.9\textwidth] {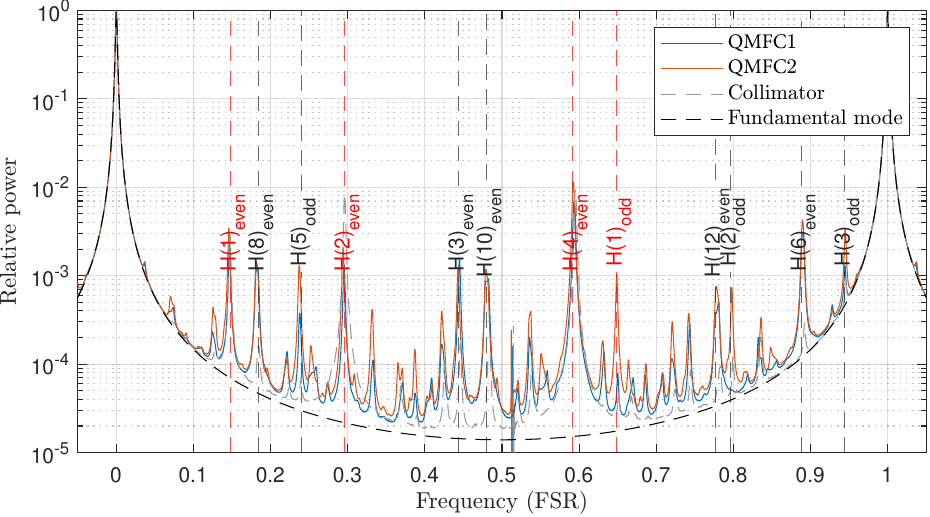}
    \caption{A mode scan over one free spectral range of the \ac{DBB} when the beam is injected by two of the \ac{QMFC} produced here and a commercially available collimator as a comparison. It is found that only one mode has a power contribution of more the $1\%$, but as this is common to all three measurements, it is likely a result of mode matching error into the resonator or a deformation on one of the mirrors. }
    \label{fig:mode_Scan}
\end{figure}
\par
The higher order mode content was studied using the optical resonator in the \ac{DBB}. The transmitted power was then measured as the cavity length was swept over one free spectral range. This was done 15 times and averaged over the result. With a linear optical resonator, the modes with the same total power would resonate together, but the \ac{DBB} used a triangular resonator to break this degeneracy so that modes with an odd or even distribution can be distinguished. We define an even mode as one where when discussing a mode or HG(m,n), m is even, and an odd mode as when m is odd.  Sample 1 and 2 were tested, as well as a commercially available collimator as a comparison. The results are shown in Figure \ref{fig:mode_Scan}.
\par
It was found that \ac{QMFC} 1 had a total higher order mode content of $2.1\pm0.05\%$, while \ac{QMFC} 2 had $3.4\pm0.05\%$, and the commercial collimator $2.2\pm0.1\%$. The first \ac{QMFC} has a comparable beam purity to the commercial piece, while the second had slightly higher mode content. This difference is largely driven by the Hermite Gauss (0,1) mode, which could easily arise from a small deformation on the fiber tip, a small misalignment of the beam to the DBB, or dust particle on the lens of the collimator. Ultimately, it is found that the \acp{QMFC} produce beams of sufficient mode purity, but if an exceptionally high mode purity is required, redundant samples should be planned to circumvent the chance of contaminants spoiling some fraction of them. It should, however, be noted that all these measurements are upper estimates of the total loss as they also include all losses in the fundamental mode in the resonator itself due to misalignment of the beam with the resonator.
\subsection{Pointing stability}
Another key measure of the \ac{QMFC} is that the resulting beam is not jittering around its axis. We require the beam to have good pointing stability to maintain alignment of beams with optical experiments, to prevent additional beam pointing noise in length measurements. To measure this, the \ac{DBB} was again used, using the reflected port of the optical resonator. The resonator's length was stabilised to the laser frequency on the fundamental mode of the resonator. The light from the reflected port is measured using two \acp{QPD}, which construct a Differential Wavefront Sensing (DWS) to feedback control the alignment of the incoupling mirrors of the cavity. Using the calibrated error signal and feedback signal, the beam tilt can be extracted. A thorough discussion of the method is detailed in \cite{Kwee2007}.
\begin{figure}
    \centering
    \subfigure[Pitch]{\includegraphics[width=.49\textwidth] {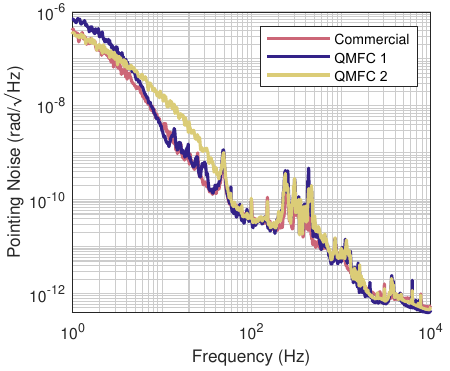}}
     \subfigure[Yaw]{\includegraphics[width=.49\textwidth] {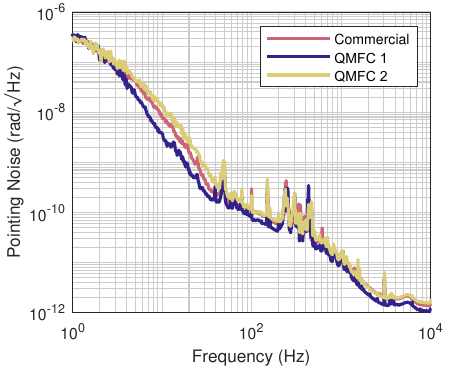}}
    \caption{The pointing stability of two \ac{QMFC} and a commercial collimator as measured by the \ac{DBB} using a DWS method. The FIOS produced here matches or exceeds the performance of the commercial coupler and likely are at the noise floor of the experiment as a whole. }
    \label{fig:beamPoiting}
\end{figure}
\par
The resulting measurements on the beam pointing of two \ac{QMFC} and a commercial coupler are shown in Figure \ref{fig:beamPoiting}. Here we see that \ac{QMFC} 1 matched or beat the performance of the commercial coupler across all frequencies in pitch and yaw, while \ac{QMFC} 2 showed some extra pitch motion at low frequencies. It is difficult to determine whether this is directly a result of the \ac{QMFC} itself or rather the adaptor mount to raise it to the correct beam profile; hence, this acts as an upper estimate of the beam pointing of the sample. Still, the results from \ac{QMFC} 1 show that, at the very least, we would be limited by mounts and posts for the non-bonded case. 
\subsection{Back Reflection}
The QMFC has several interfaces where a potential back reflection can occur. Such as in the bond between the lens and cuboid and at the fiber weld. We, therefore, wished to measure how much power was reflected back into the fiber and ultimately towards the laser. With high-power applications, this back reflection can be problematic, and so we wish to estimate this effect.
The back reflection was measured using a fiber circulator. Approximately 70\ mW of optical power was injected into the collimator, and the ratio of power in the back reflected port of the circulator to that transmitted through the collimator was measured.  This tells us only the amount of light directly reflected into the fibre and not the amount scattered, which is otherwise difficult to distinguish from optical losses due to fibre coupling. A coupling efficiency to the QMFC of above 90$\%$ was shown in each piece. The setup is then limited by the isolation of the ports of the circulator, which is 50\,dB. Tests on a commercial fiber collimator showed a back reflection ratio of 44\,dB. 
\par
Each of the 8 samples was tested in this setup. It was noticed that there was a significantly larger back reflection of about 6\,dB from silicate bonded samples than those that were optically contacted. The optically contacted samples showed a back reflection ratio of ($36.4\pm1.4$)\,dB, while the silicate bonded ones showed a back reflection ratio of ($29.2\pm1.0$)\,dB. Both these results are considerably higher than those of the commercial fiber collimator. The amount of back-reflected light is of concern when high-power applications are used. In this case, optical contacting is likely the better choice of bonding method, but still, the high back reflected power would warrant additional isolation stages such as Faraday isolators.

\section{Discussion of Optical Head}
\label{sec:disc}
The mode matching, mode purity, beam pointing, and back reflection have been validated to at least match a standard fibre collimator while maintaining the benefit that it is easy to bond any of the flat surfaces to, for example, a flat baseplate used as a payload demonstration.  
\par
A specific beam profile within tolerances can be achieved using simple analytical modelling to define the length and tolerance of the cuboid. When a stricter tolerance on beam profile is needed, a laser ablation or precision polishing manufacturing method can be deployed to achieve the required length tolerances. Some of the assemblies showed some astigmatism, this likely occurred due to the bonding of the lens with unequal stress across the whole surface or small misalignments of the fiber weld. 
\par
We have shown that neither the fiber weld nor the optical contact significantly degrades the profiles resulting beam when compared to commercially available, non-bonded fiber collimators, with $\rm{M}^2$ of 1.125 having been demonstrated. This means the laser beams from these are of have low enough higher order mode content for use in precision sensors that need well-defined beams, such as optical resonators or DWS sensing. 
\section{Conclusions}
\label{sec:conc}
We have demonstrated a means of achieving a simple-to-build optical head that can be used to inject collimated laser light into experiments. The design is well suited to the many experiments that wish to test interferometric setups, but need ultra low expansion glass baseplates to minimise thermal drifts. Many experiments could utilize similar or adapted designs to the QMFC shown here instead of injecting light from commercial fiber collimators mounted off the glass baseplate, which can otherwise limit the performance of such experiments. Our Quasi Monolithic Fiber Collimator design can be used on hardware demonstrations to allow for quicker iteration of the potential payload of physics experiments without needing to assemble a full, time-consuming and complex to manufacture \ac{FIOS} for each iteration.    
\section*{Acknowledgements}
This work was funded  in the framework of the Max-Planck-Fraunhofer Kooperationsprojekts ``Optomechaniken hoher G\"{u}te f\"{u}r quantenrauschliterte Gravitationalwellendetektion (HighQG)" \& ``Glass Technologies for the Einstein Telescope (GT4ET)".


\begin{thebibliography}{10}
\newcommand{\enquote}[1]{``#1''}

\bibitem{Killow2016}
C.~J. Killow, E.~D. Fitzsimons, M.~Perreur-Lloyd, D.~I. Robertson, H.~Ward, and
  J.~Bogenstahl, \enquote{Optical fiber couplers for precision spaceborne
  metrology,} {\protect\JournalTitle{Applied Optics}} \textbf{55}, 2724 (2016).

\bibitem{Chwalla2016}
M.~Chwalla, K.~Danzmann, G.~F. Barranco, E.~Fitzsimons, O.~Gerberding,
  G.~Heinzel, C.~J. Killow, M.~Lieser, M.~Perreur-Lloyd, D.~I. Robertson,
  S.~Schuster, T.~S. Schwarze, M.~Tröbs, H.~Ward, and M.~Zwetz,
  \enquote{Design and construction of an optical test bed for {LISA} imaging
  systems and tilt-to-length coupling,} {\protect\JournalTitle{Classical and
  Quantum Gravity}} \textbf{33}, 245015 (2016).

\bibitem{Veggel2014}
A.-M.~A. van Veggel and C.~J. Killow, \enquote{Hydroxide catalysis bonding for
  astronomical instruments,} {\protect\JournalTitle{Advanced Optical
  Technologies}} \textbf{3}, 293--307 (2014).

\bibitem{Bogenstahl2017}
J.~Bogenstahl, C.~Diekmann, E.~D. Fitzsimons, R.~Fleddermann, E.~Granova, C.~J.
  Killow, J.~Pijnenburg, D.~I. Robertson, A.~Shoda, A.~Sohmer, M.~Tröbs,
  H.~Ward, D.~Weise, L.~d’Arcio, M.~Dehne, G.~Heinzel, H.~Hogenhuis,
  M.~Perreur-Lloyd, A.~Taylor, and G.~Wanner, \enquote{Optical bench
  development for lisa,} in \emph{International Conference on Space Optics —
  ICSO 2010,}  N.~Kadowaki, ed. (SPIE, 2017).

\bibitem{Danzmann2017}
K.~Danzmann, M.~Dehne, O.~Gerberding, D.~Schütze, G.~Stede, R.~Spero,
  D.~Shaddock, G.~Heinzel, B.~Sheard, N.~Brause, C.~Mahrdt, V.~Müller,
  W.~Klipstein, W.~M. Folkner, K.~Nicklaus, and P.~Gath, \enquote{Laser ranging
  interferometer for grace follow-on,} in \emph{International Conference on
  Space Optics — ICSO 2012,}  E.~Armandillo, N.~Karafolas, and B.~Cugny, eds.
  (SPIE, 2017).

\bibitem{nicklaus2017}
K.~Nicklaus, M.~Herding, A.~Baatzsch, M.~Dehne, C.~Diekmann, K.~Voss,
  F.~Gilles, B.~Guenther, B.~Zender, S.~Boehme \emph{et~al.}, \enquote{Optical
  bench of the laser ranging interferometer on grace follow-on,} in
  \emph{International Conference on Space OpticsICSO 2014,}  (2017), p.
  105632I.

\bibitem{Boehme2009}
S.~Boehme, E.~Beckert, R.~Eberhardt, and A.~Tuennermann, \enquote{Laser
  splicing of end caps: process requirements in high power laser applications,}
  in \emph{Laser-based Micro- and Nanopackaging and Assembly III,}
  W.~Pfleging, Y.~Lu, K.~Washio, W.~Hoving, and J.~Amako, eds. (SPIE, 2009).

\bibitem{Boehme2017}
S.~Böhme, S.~Fabian, A.~Kamm, T.~Peschel, A.~Tünnermann, M.~Dehne,
  E.~Beckert, and K.~Nicklaus, \enquote{High reproducible co2 laser spliced
  fiber-collimator for a space borne laser system,} in \emph{International
  Conference on Space Optics — ICSO 2016,}  N.~Karafolas, B.~Cugny, and
  Z.~Sodnik, eds. (SPIE, 2017).

\bibitem{Pierangelo2016}
C.~Pierangelo, B.~Millet, F.~Esteve, M.~Alpers, G.~Ehret, P.~Flamant,
  S.~Berthier, F.~Gibert, O.~Chomette, D.~Edouart, C.~Deniel, P.~Bousquet, and
  F.~Chevallier, \enquote{Merlin (methane remote sensing lidar mission): an
  overview,} {\protect\JournalTitle{EPJ Web of Conferences}} \textbf{119},
  26001 (2016).

\bibitem{Kwee2007}
P.~Kwee, F.~Seifert, B.~Willke, and K.~Danzmann, \enquote{Laser beam quality
  and pointing measurement with an optical resonator,}
  {\protect\JournalTitle{Review of Scientific Instruments}} \textbf{78} (2007).

\end{thebibliography}
\end{document}